\documentclass[]{aastex}

\usepackage{xcolor}

\fullcollaborationName{The Friends of AASTeX Collaboration}

\begin{document}

\title{On the shape of SEP electron spectra: The role of interplanetary transport}


\author{R.D. Strauss\altaffilmark{1}}
\affil{Center for Space Research, North-West University, Potchefstroom, 2522, South Africa}

\and
\author{N. Dresing\altaffilmark{2},  A. Kollhoff \altaffilmark{3}, \and M. Br\"udern \altaffilmark{4}}
\affil{Institut f\"ur Experimentelle und Angewandte Physik, Universit\"at Kiel, 24118, Kiel, Germany}


\altaffiltext{1}{dutoit.strauss@nwu.ac.za} 
\altaffiltext{2}{dresing@physik.uni-kiel.de}
\altaffiltext{3}{kollhoff@physik.uni-kiel.de}
\altaffiltext{4}{bruedern@physik.uni-kiel.de}

\begin{abstract}

We address the effect of particle scattering on the energy spectra of solar energetic electron events using i) an  observational and ii) a modeling approach. i) We statistically study observations of the STEREO spacecraft making use of directional electron measurements made with the SEPT instrument in the range of 45 -- 425 keV. We compare the energy spectra of the anti-sunward propagating beam with that one of the backward scattered population and find that, on average, the backward scattered population shows a harder spectrum with the effect being stronger at higher energies. ii) We use a numerical SEP transport model to simulate the effect of particle scattering (both in terms of pitch-angle and perpendicular to the mean field) on the spectrum. We find that pitch-angle scattering can lead to spectral changes at higher energies (E $>100$ keV) and further away from the Sun (r $> 1$ au) which are also often observed. At lower energies, and closer to the Sun the effect of pitch-angle scattering is much reduced so that the simulated energy spectra still resemble the injected power-law functions. When examining pitch-angle dependent spectra, we find, in agreement with the observational results, that the spectra of the backward propagating electrons are harder than that of the forward (from the Sun) propagating population. {We conclude that {\it Solar Orbiter} and {\it Parker Solar Probe} will be able to observe the unmodulated omni-directional SEP electron spectrum close to the Sun at higher energies, giving a direct indication of the accelerated spectrum. }
\end{abstract}


\section{Introduction}

Solar energetic particle (SEP) spectra, observed in-situ at e.g. 1 au, can provide a wealth of information regarding the processes responsible for their acceleration \citep[see, for example, the results and discussion by][]{brunoetal2018}. However, interplanetary transport may significantly alter the observed spectral shape during propagation, making it much more difficult to disentangle acceleration and transport effects. In this work, we address this issue by analyzing observations of SEP electrons and comparing these to transport simulations in the inner heliosphere, where we focus on the resulting spectral shape.\\

It is generally believed that SEP events can be classified as either gradual (also referred to as proton-rich) or impulsive (also referred to as electron-rich) and that each group has a different accelerator, with acceleration in solar flares generally thought to be responsible for electron events \citep[][]{reames1999,reames2013}. This assumption can be tested by comparing observed in-situ electron spectra at e.g. 1 au with remote sensing observations of the same flaring region \citep[see e.g.][]{kruckeretal2007}. However, electron transport effects, if any, must be accounted for, and this work will attempt to characterize such effects.\\

Recently, \citet{Dresing2020} analyzed the energy spectra of all solar energetic electron events observed by the two STEREO spacecraft since its mission began in 2007. The STEREO Solar Electron and Proton Telescope \citep[SEPT,][]{Muller-mellin2008} measures electrons in the energy range between 45 and 425\,keV. Of the total 781 events, nearly half showed spectra with a broken-power-law shape as previously reported in similar energy ranges especially for impulsive solar energetic electron events \citep[e.g. ][]{kruckeretal2009}. However, the mean break energy found by \citet{Dresing2020} was about 120\,keV, which is much higher compared to results by \citet{kruckeretal2009} who found a mean break energy of 60\,keV using Wind/3DP data. However, \citet{Dresing2020} note that the energy range of SEPT together with their fitting procedure does not allow to determine a spectral break below 70\,keV. It it therefore not known if the spectral breaks at higher energies are additional spectral breaks or if the break energy has significantly changed. The term {\it spectral break} might be misleading in this context, as it implies a sudden and discontinuous change in the power-law index. These {\it spectral changes/transitions} are, however, relatively smooth as illustrated in Appendix \ref{Sec:sharpness}. \\

Wave-particle interactions which are responsible for Langmuir- and radio-wave generation \citep[e.g.][]{kontarreid2009} can lead to low energy electrons suffering significant energy losses.  Transport modeling by \citet{kontarreid2009} shows that this effect can lead to the formation of a spectral change between 4 and 80\,keV depending on the model parameters. These authors therefore argue that the spectral changes observed at energies of about 60\,keV are caused by this wave-generation effect. The spectral break energies found by \citet{Dresing2020} did not show any dependence on the observed peak intensities of the events, as would be expected in case of wave-generation effects \citep{kontarreid2009}, and this absence suggests another process is responsible for these spectral breaks. On the one hand, this could be the acceleration mechanism itself, while, on the other hand, pitch-angle diffusion, being a process that affects higher energy electrons, may be another candidate. \\

\begin{figure*}
\begin{center}
\includegraphics[width=130mm]{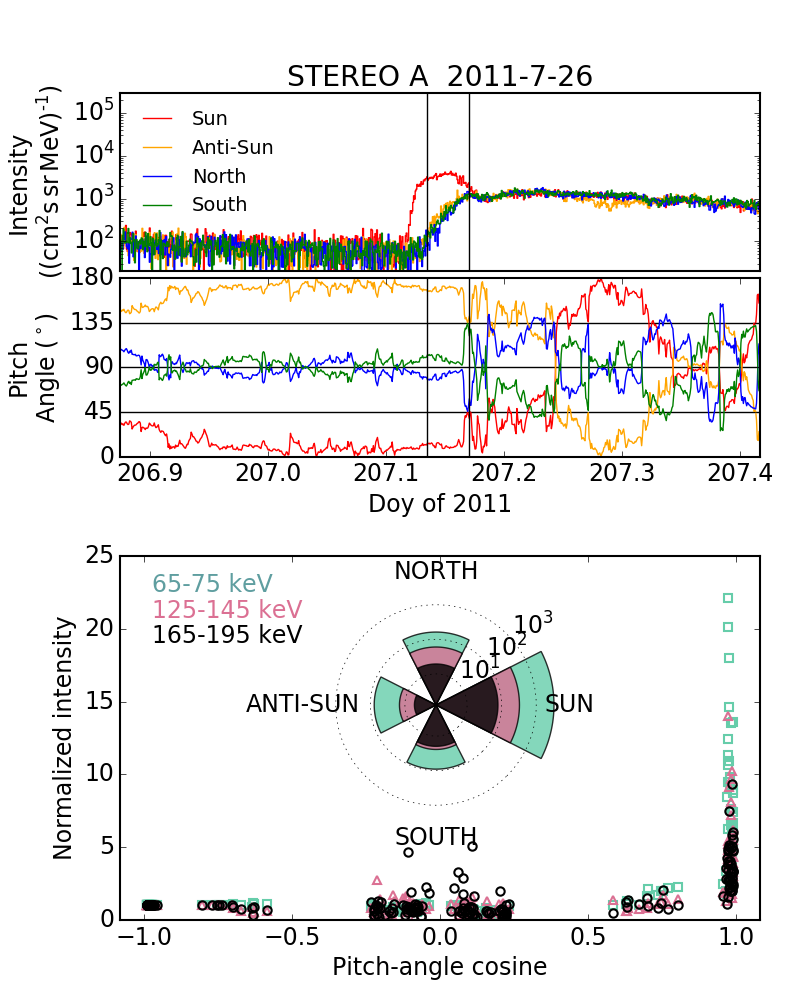}
\caption{STEREO~A/SEPT observations of an SEP event on 2011 July 26. The two top panels show the intensity of 65--75 keV electrons in the four viewing directions of the instrument (top) and the corresponding pitch-angles (middle). The bottom panel shows the pitch-angle distributions for the time period marked by vertical lines in the upper two panels in three different energy channels (indicated on the legend). Each pitch-angle distribution is normalized to the intensity in the pitch-angle bin close to pitch-angle cosine $\mu \approx -1$. The inset in the bottom panel shows the intensities measured in the four viewing directions of SEPT represented by sectors during one timestamp at the beginning of the period. The sector widths correspond to the opening angles of the telescopes and the colors mark the different energy channels. Note the logarithmic radial axis used for this polar plot. \label{fig:en_dep_pad}}
\end{center}
\end{figure*}

\citet{droge2000} showed that the pitch-angle diffusion coefficient, which describes the scattering of energetic particles at magnetic irregularities, is energy dependent both for electrons and protons, although different for the two species. When interplanetary transport modeling is applied to SEP observations at different energies, this leads to changing mean free paths \citep[see also][]{Agueda2014}, which in the case of near-relativistic electrons, decreases with energy. This effect can also be seen in the pitch-angle distributions of SEP events observed at 1 au which become increasingly anisotropic at lower energies \citep[e.g., see Fig. 2 of ][]{tan2009}. Fig. \ref{fig:en_dep_pad} shows exactly such observations by STEREO~A/SEPT on 2012 May 26. The top panel shows the intensity of 65--75 keV electrons in the four viewing directions of the instrument with the corresponding pitch-angles in the panel below. The event shows a strong anisotropy with a much higher intensity in the sunward-pointing telescope. The bottom panel shows normalized pitch-angle distributions in three different energy channels (as indicated in the legend) for the time period marked by vertical lines in the upper panels. The lowest energy bin clearly shows the strongest anisotropy, i.e. the largest intensity difference between the anti-sunward propagating particle beam (seen close to pitch-angle cosine $\mu \approx 1$) and the opposite direction ($\mu \approx -1$). The highest energy bin shows the smallest anisotropy which is expected due to the stronger scattering of the higher energy electrons. The inset shows a snapshot of the measured intensities in the four different viewing cones of SEPT for one selected timestamp. In this polar representation the radial axis represents the intensity (note the logarithmic scale) and the widths of the sectors represent the opening angles of the four viewing directions of SEPT. The energy dependence of the intensity differences between the different viewing directions is clearly visible.\\

If the scattering conditions for solar energetic electrons are energy dependent, this can result in a significant change of the energy spectrum during the transport of the particles from the Sun to the observer, and therefore hamper the interpretation of the spectrum with respect to its underlying acceleration process. 
{red}{See \citet{LiLee2015} for an analysis of the effect of scatter-dominated interplanetary transport on the spectra of large gradual proton events.}
In the following part of this paper (section \ref{sec:obs}) we aim at quantifying this effect using data from STEREO/SEPT. In the second part (section \ref{sec:model}) we will make use of the spatially 2D SEP transport model, discussed by \citet{straussetal2017}, to simulate transport effects on electron spectra at different radial positions and at different levels of magnetic connectivity to the source region.\\

In contrast to the previous work by \citet{kontarreid2009}, who primarily focused on effects taking place only close to the Sun and affecting the low-energy (primarily below $\sim 10-50$ keV) part of the electron spectra, we neglect these low energy-loss processes and focus on the higher energy regime.

\begin{figure*}
\begin{center}
\includegraphics[width=100mm]{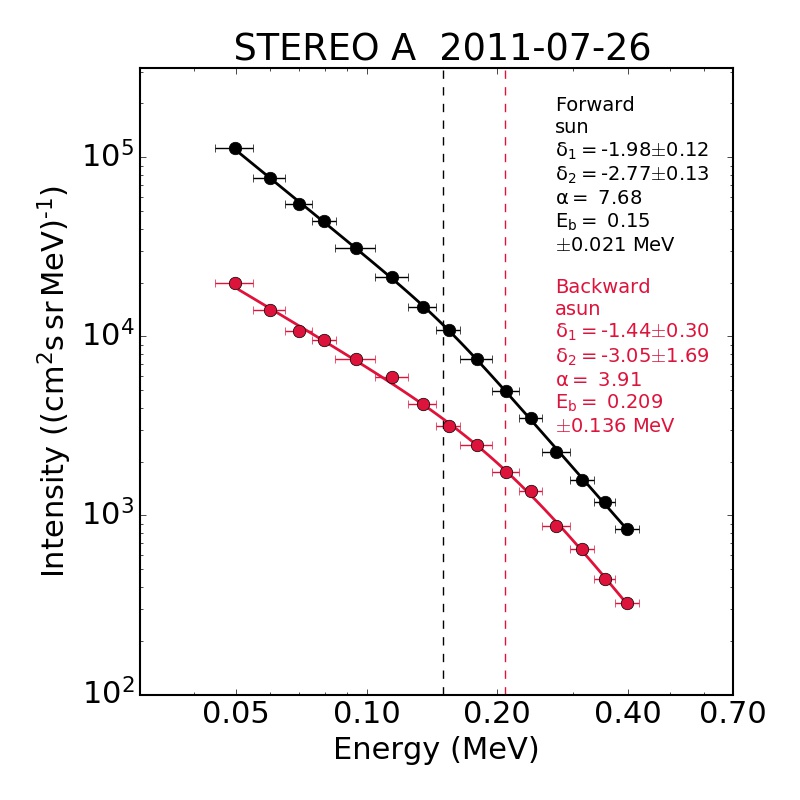}
\caption{Example TOM spectra observed with STEREO~A/SEPT with black representing the forward direction, observed in the Sun telescope, and red the backward direction (Anti-Sun telescope), respectively. The vertical dashed lines show the position of the respective break energies. \label{fig:obs_example}}
\end{center}
\end{figure*}

\begin{figure*}
\begin{center}
\includegraphics[width=0.48\textwidth]{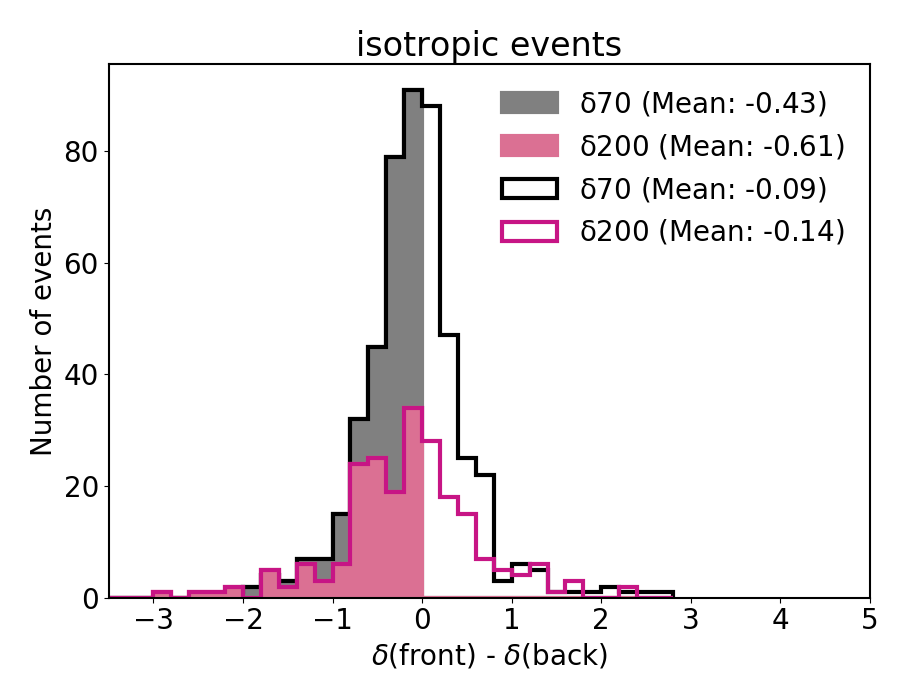}
\includegraphics[width=0.48\textwidth]{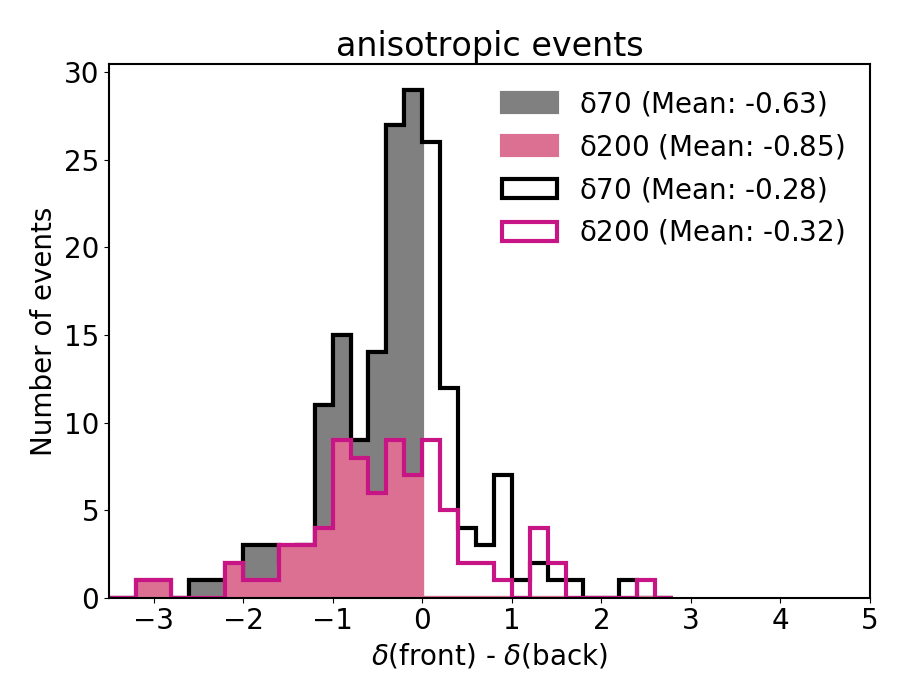}
\caption{Difference of spectral indices between forward and backward directions (within each event) for the spectral index at 70 and 200 keV, respectively. Right: strongly anisotropic events, left: all other events. \label{fig:obs_diff_hist_gamma70_200}}
\end{center}
\end{figure*}

\section{Observations}\label{sec:obs}

To analyze the effect of interplanetary pitch-angle scattering on the energy spectrum of an SEP event one would ideally utilize multi-spacecraft observations of the same event. While the radial effect could be analyzed through magnetically aligned spacecraft, situated at different radial distances, the longitudinal effect, probably caused by perpendicular transport, would require multi-spacecraft observations by longitudinally separated observers. Furthermore, in the latter case one would have to assume that the SEP injection, which might be spatially extended, does not have a longitudinal dependence. Given the current number of available spacecraft it is nearly impossible to guarantee this, especially in the case of extended injection regions. Moreover, the number of multi-spacecraft events observed by magnetically-aligned observers at different radial distances is very limited. In this work we chose another approach to determine the imprint of transport effects on the energy spectrum which makes use of single spacecraft observations in different viewing directions and therefore can provide good event statistics. For this purpose we determine and compare the energy spectrum of each event in two anti-parallel viewing directions: one observing the SEP distribution directly streaming away from the Sun ('forward direction'), and the other observing the SEP distribution which was scattered back, leading to the particles entering the instrument from the 'backward direction' and therefore showing lower peak intensities than the anti-sunward propagating distribution. This allows us to compare two SEP distributions which experienced different amounts of scattering but are related to the same event. We make use of the events studied by \citet{Dresing2020} who analyzed 781 time-of-maximum (TOM) energy spectra of the anti-sunward propagating portion of the solar energetic electron events (forward direction) observed by the SEPT instruments aboard the two STEREO spacecraft providing a good statistical sample.\\ 

Each of the four viewing directions of STEREO/SEPT measures near-relativistic electrons in 15 energy channels in the range from $45-425$ keV. Two of them are aligned with the nominal \citet{parker1958} spiral in the ecliptic plane viewing toward the Sun ('Sun') and away from the Sun ('Anti-Sun'), and the other two telescopes are looking perpendicular to these toward north ('North') and south ('South'). As the telescope which observes the anti-sunward propagating SEP beam (forward telescope) may vary according to the interplanetary magnetic field configuration, we chose the backward telescope accordingly as the one viewing into the anti-parallel direction compared to the forward telescope. However, if a clear bi-directional distribution was observed, potentially caused by an efficient reflecting boundary or the presence of a magnetic cloud and therefore not dominated by scattering, we use one of the other two telescopes, perpendicularly orientated with respect to the forward telescope, as the 'backward telescope'.\\

To be able to observe the effect of interplanetary transport on the energy spectra in these single-spacecraft data, we only select those events showing significant anisotropies close to the peak intensity time (within 60 minutes). In the case of isotropic events, the different viewing directions show nearly the same intensity-time profile and therefore also the same spectra. \citet{Dresing2020} have reported different spectral shapes, i.e. broken or single power-law functions, both occurring in their large statistical sample. In this work we do not strictly assume a single or a broken-power-law to describe the observed spectrum but we also exploit broken-power-law functions with a smooth transition or break (Eq. \ref{Eq:broken_pl}) and single-power-law functions bending over into an exponential cutoff  (Eq. \ref{Eq:pl_exp})

\begin{equation}
I(E) = I_0\left( \frac{E}{E_0}\right)^{\delta_1} \left( \frac{E^\alpha + E_b^\alpha}{E_0^\alpha + E_b^\alpha}\right)^{(\delta_2-\delta_1)/\alpha}
\label{Eq:broken_pl}
\end{equation}

\begin{equation}
I(E) = I_0 \left( \frac{E}{E_0}\right)^{\delta} \cdot e^{-(E/E_c)^2}
\label{Eq:pl_exp}
\end{equation}

{In Eq. \ref{Eq:broken_pl} $\delta_1$ and $\delta_2$ are the spectral indices below and above the break energy $E_b$, respectively and $\alpha$ describes the sharpness of the break with larger values corresponding to sharper breaks. $E_0$ is a reference energy. Eq. \ref{Eq:pl_exp} obviously describes only a single-power-law with spectral index $\delta$ which goes over into an exponential decay at a cutoff energy $E_c$. The reader is refereed to Appendix \ref{Sec:sharpness} for a more detailed discussion on the spectral shapes.}\\
\begin{figure*}
\begin{center}
\includegraphics[width=0.48\textwidth]{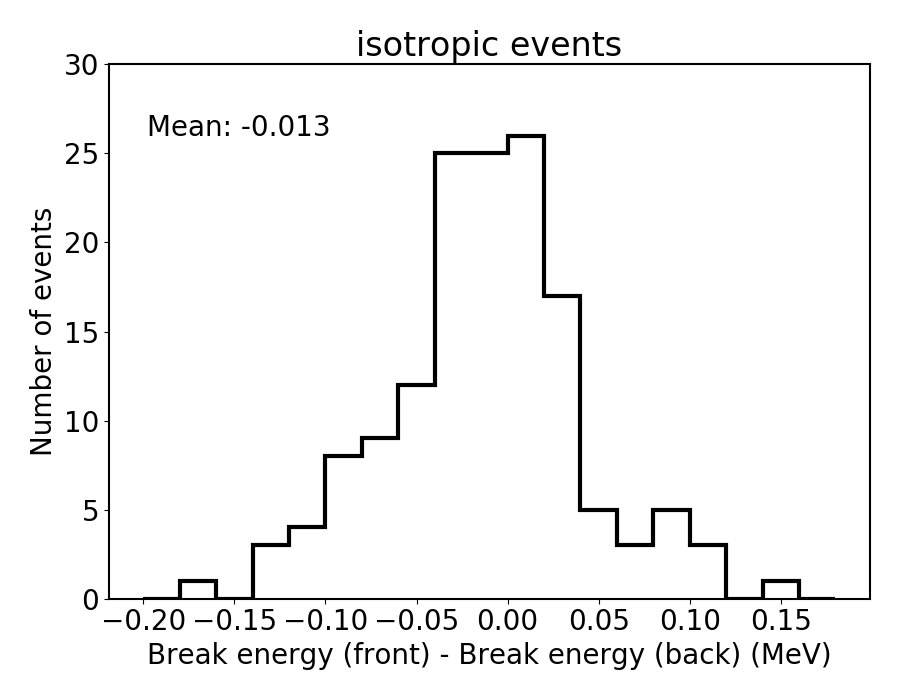}
\includegraphics[width=0.48\textwidth]{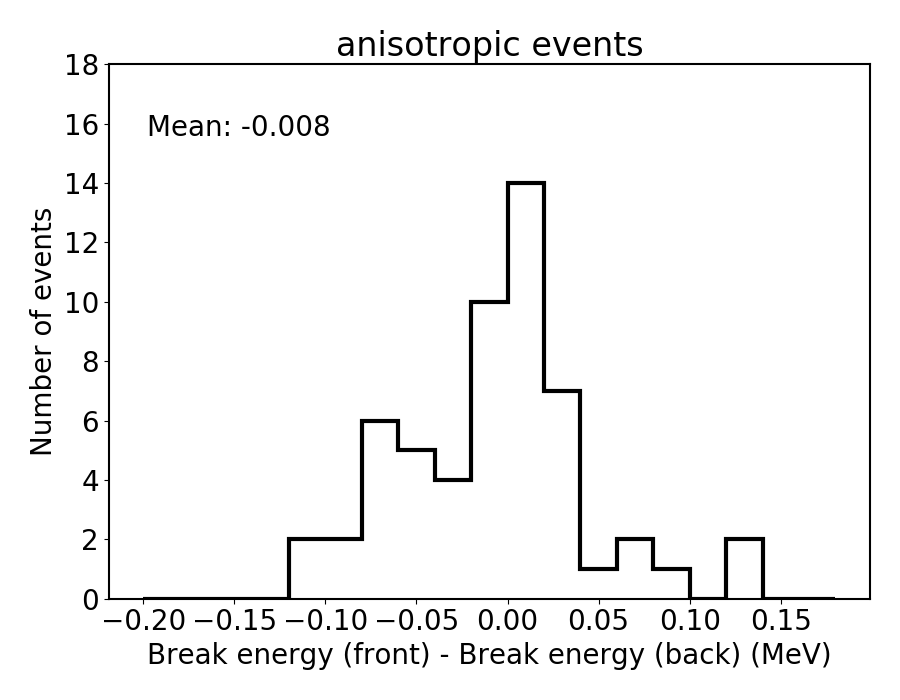}
\caption{Distribution of the break energy change between forward and backward directions. Only events showing a broken-power-law in both forward and backward directions are considered here. Right: strongly anisotropic events, left: all other events. \label{fig:obs_diff_break}}
\end{center}
\end{figure*}
Fig. \ref{fig:obs_example} shows forward (black) and backward (red) spectra for the event shown in Fig. \ref{fig:en_dep_pad}. Both spectra show broken-power-laws, however with rather round transitions or breaks. The overall intensities of the backward spectrum are lower {red}{and it represents a harder spectrum
compared to the forward one.} Despite a large event-to-event variation this example event represents the average behaviour observed in our statistical sample. However, we observe different spectral shapes in our sample (i.e., single-power-laws, broken-power-laws with variable sharpness of the break, and power-laws with exponential cutoffs), and these often change from the forward to the backward spectrum. In order to analyze all events in our sample together, regardless of their spectral shape or the position of the eventual break point $E_b$, we define two spectral indices $\delta70$ and $\delta200$ corresponding to the spectral index found in the energy range around 70\,keV and 200\,keV, respectively. In the case of single-power-law events, $\delta70$ and $\delta200$ have equal values. For double-power-law events, defined by the two spectral indices $\delta_1$ and  $\delta_2$, we define $\delta70$ and $\delta200$ as follows:
\begin{eqnarray}
\delta70 &=& \left\{ \begin{array}{ccc}
     \delta_1 \quad & :E_b & > 70 \mathrm{keV} \\
     \delta_2 \quad & :E_b & < 70 \mathrm{keV} \end{array} \right. \\
\delta200 &=& \left\{ \begin{array}{ccc}
     \delta_1 \quad & :E_b & > 200 \mathrm{keV} \\
     \delta_2 \quad & :E_b & < 200 \mathrm{keV} \end{array} \right. 
\end{eqnarray}

Fig. \ref{fig:obs_diff_hist_gamma70_200} shows the distributions of spectral changes from forward to backward direction (within each event) at 70 keV ($\delta70$, pink) and at 200 keV ($\delta200$, black). The statistics for the $\delta70$ distributions are larger because many events do not extend to $\geq$200 keV or this energy is $\geq E_c$ in case the fit found only a single-power-law with exponential cutoff. While the right hand figure shows the distributions only for events showing strong anisotropies, the left hand figure shows the distributions for the rest of the events which are nearly isotropic. The mean values (provided in the figure legends) of all distributions are negative, representing, on average, a spectral hardening from forward to backward direction, similar to the example in Fig. \ref{fig:obs_example}. The spectral hardening is clearly stronger for the anisotropic sample as expected due to isotropic fluxes not showing the effect (see the discussion above). Only the filled parts of the distributions correspond to a spectral hardening from forward to backward direction and their mean values are provided separately with the mean values of the whole distributions in the figure legend. The more negative mean values of the $\delta200$ distributions seem to suggest a stronger average spectral hardening effect at higher energies.
The reason for the occasional spectral softening from forward to backward spectrum, i.e. the positive parts of the distributions, is not clear.
However, time-extended and possibly also energy-dependent injection functions at the Sun could cause this effect.\\


Fig. \ref{fig:obs_diff_break} shows the distribution of the changes in the break energy from forward to backward spectra (only for those events showing a broken-power-law in both directions). Both the distributions of the anisotropic (right) and the isotropic (left) events do not show a systematic change but a wide distribution of break-energy changes up to about 100\,keV towards higher and lower energies is observed. \\

%

In the following sections we will attempt to explain the changing forms of the SEP electron spectra by simulating these changes with a numerical SEP transport model. It is important to note that we do not aim to fit each individually observed spectrum, but we rather aim to simulate their general, or average, properties. It is well known that the transport parameters can be very different between SEP events \citep[e.g.][]{pachecoetal2019}, leading to large inter-event variability that cannot be reproduced with a model using a single set of transport parameters. Later in this paper we also speculate on a possible explanation for the large inter-event variability that is observed {by varying the assumed transport parameters in the SEP model}.

\section{Transport Model}\label{sec:model}

\begin{figure*}
\begin{center}
\includegraphics[width=75mm]{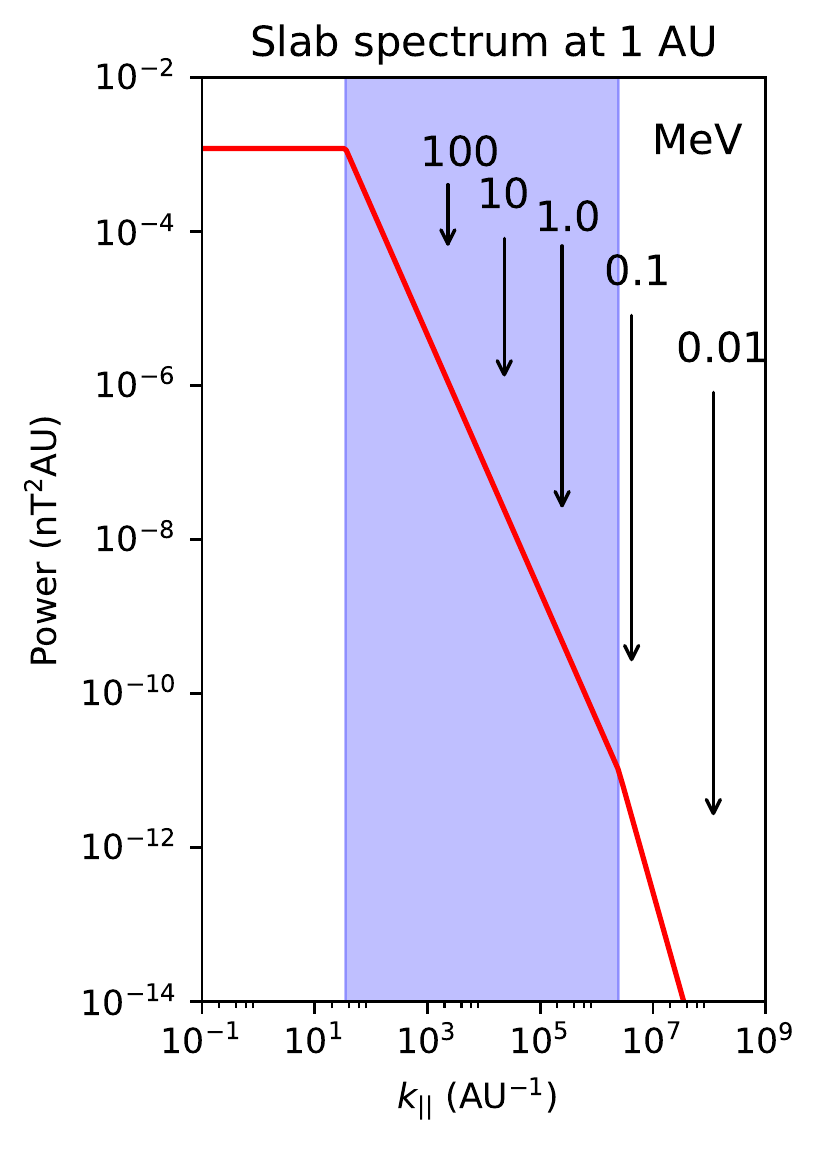}
\includegraphics[width=75mm]{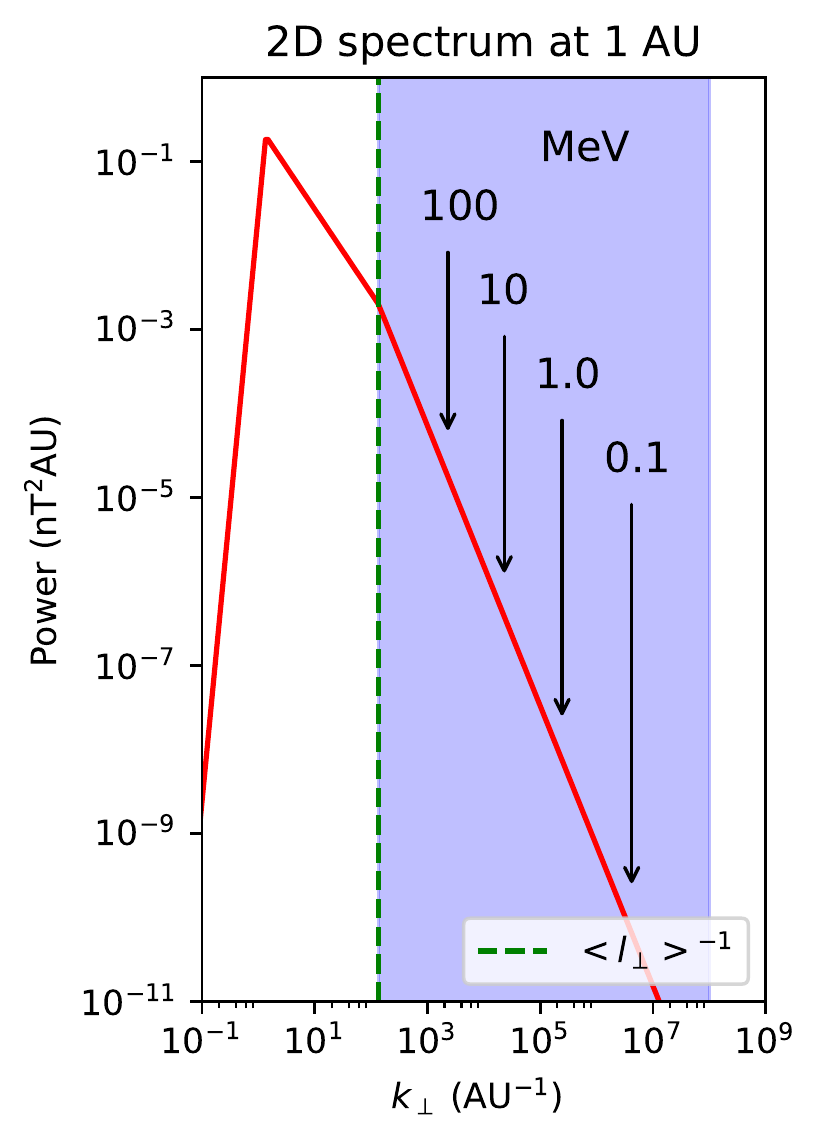}
\caption{The turbulence power spectra assumed in this work. The vertical arrows indicate the approximate wavenumbers where electrons of different energies could potentially resonate with wave-like turbulent features.  \label{fig:spectra_resonate}}
\end{center}
\end{figure*}

\begin{figure*}
\begin{center}
\includegraphics[height=100mm]{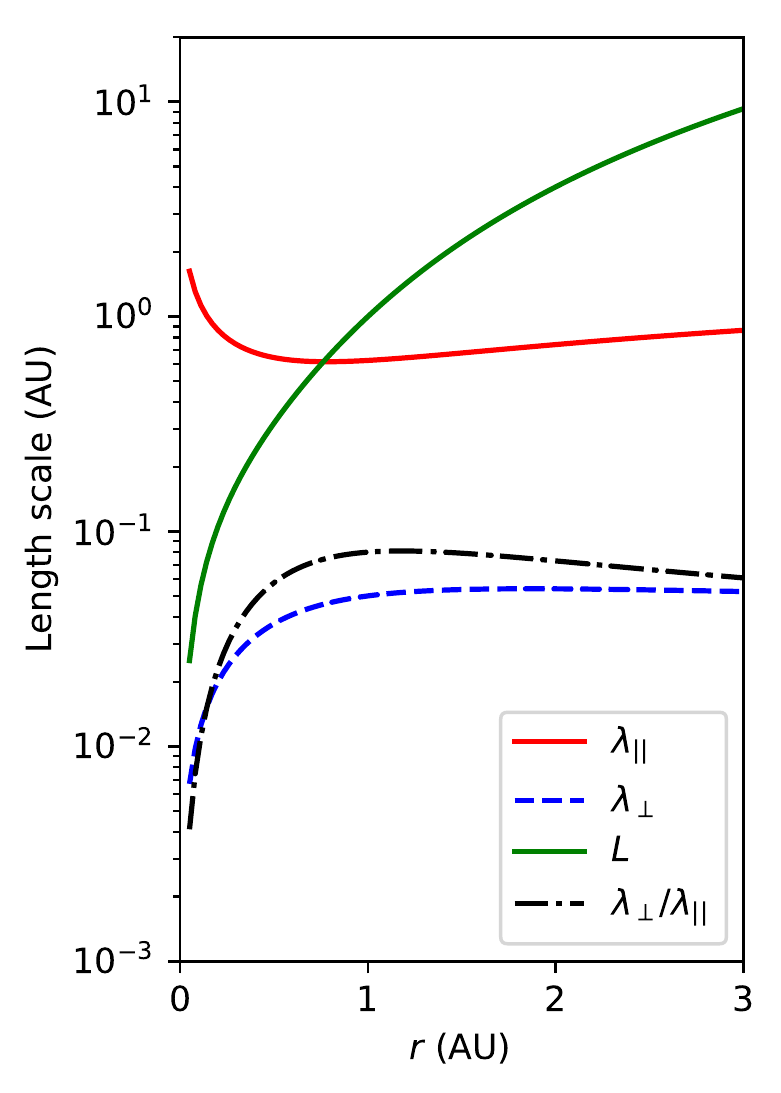}
\hspace{5mm}
\includegraphics[height=100mm]{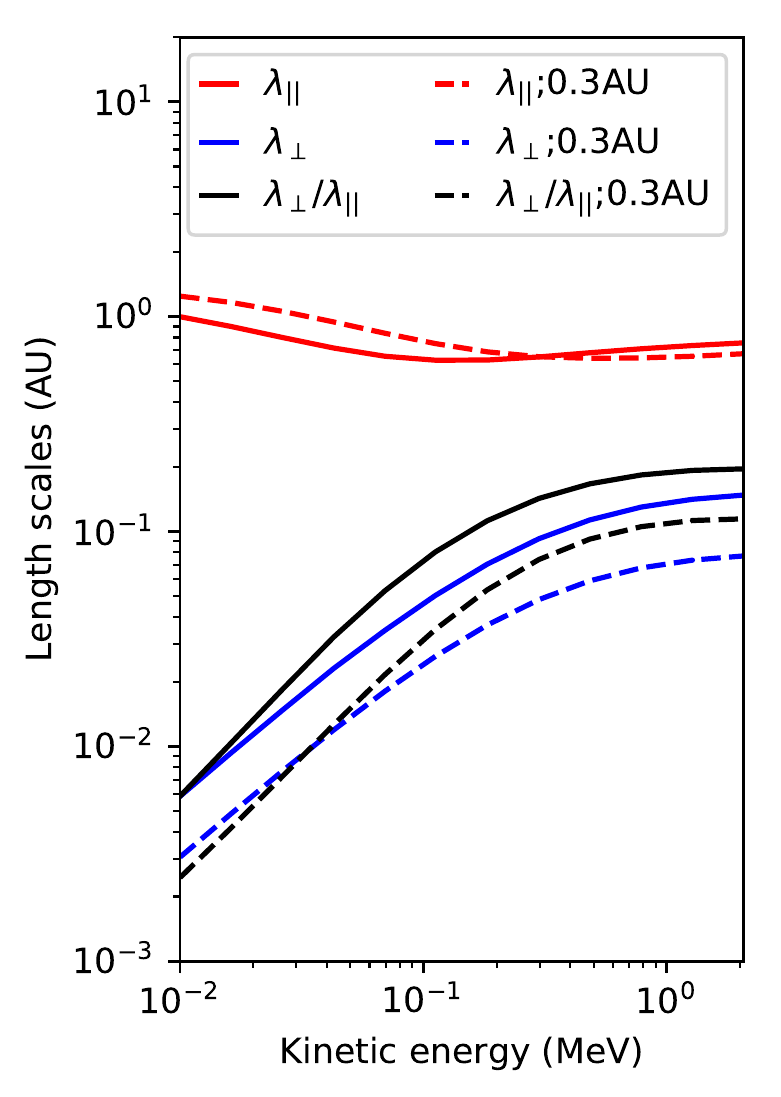}
\caption{The resulting transport length scales as a function of radial distance at 100 keV (left panel) and as a function of energy at 1 au (right panel; solid lines) and 0.3 au (right panel; dashed lines).  \label{fig:transport_coeffs}}
\end{center}
\end{figure*}

In order to simulate the propagation of electrons in the turbulent interplanetary magnetic field we make use of a simplified version of the \citet{skilling1971} equation,

\begin{eqnarray}
\label{Eq:TPE}
\frac{\partial f}{\partial t} + \nabla \cdot \left( \mu v \hat{b} f \right) + \frac{\partial}{\partial \mu} \left( \frac{1-\mu^2}{2L} vf \right) &=&   
 \frac{\partial}{\partial \mu} \left(D_{\mu\mu}  \frac{\partial f}{\partial \mu} \right)\nonumber \\  
&+&   \nabla \cdot \left( \mathbf{D}^{(x)}_{\perp}\cdot \nabla f \right) 
\end{eqnarray}

which is numerically integrated using the spatially 2D model of \citet{straussfichtner2015}.  We neglect any forms of energy losses/gains which seems to be a reasonable assumption for near-relativistic electrons. Note however that the transport coefficients, $D_{\mu \mu}$ and $D_{\perp}$, are energy dependent and the resulting energy dependent transport is illustrated in this paper. Ultimately, the level of pitch-angle and perpendicular scattering depends on the geometry and magnitude of the background solar wind turbulence. We follow our previous approach \citep[see][]{straussetal2017} and specify a slab and 2D composite turbulence geometry constrained by spacecraft observations. Fig. \ref{fig:spectra_resonate} shows the so-called slab (left panel) and 2D (right panel) turbulence energy spectra, as assumed in this work, at 1 au. The form, and level, of the spectra are consistent with the form used in previous work \citep[][]{straussetal2017} and the measurements discussed therein. For each turbulent component the wave number range corresponding to the inertial range is shaded blue. Before delving into the details of the transport coefficients, it is useful to estimate with which turbulent fluctuations energetic electrons of different energies will resonate/interact. Assuming that particles will resonate with turbulent fluctuations with roughly the same scale of the gyro-radius

\begin{equation}
k_{\mathrm{res}} \sim r_L^{-1},
\label{Eq:where_resonanted}
\end{equation}

Fig. \ref{fig:spectra_resonate} indicates, with vertical arrows, at which wave number electrons with different energies can resonate with wave-like turbulent structures. Of course, this is only a first-order estimation, and, in reality, electrons with a given energy will resonate with a range of fluctuations depending, amongst other quantities, on the pitch-angle of the particles. Considering Alfv{\'e}nic waves, with different polarities $(n = \pm 1)$, propagating in different directions along the mean magnetic field $(j = \pm 1)$, the parallel resonance wave number for slab (i.e. wave-like) turbulence can be calculated to be

\begin{equation} 
k_{||}^{\mathrm{res}} = \frac{n\Omega}{V_A - j v \mu } 
\end{equation} 

in the absence of any non-linear damping effects. Here, $\Omega$ is the particle cyclotron frequency and $V_A$ the Alfv{\'e}n speed \citep[details are given in e.g.][]{straussleroux2019}. Assuming $| v_{||} | = v | \mu | \gg V_A$ for  all pitch-angles (the classical {\it fast-particle} assumption), and calculating just the magnitude of $k_{||}^{\mathrm{res}}$ by neglecting any wave polarization and/or propagation effects, we can reduce the equation above to $| k_{||}^{\mathrm{res}} | \sim  \Omega/ | v_{||} |$. Writing $\Omega = v_{\perp} / r_L$ where $r_L$ is the maximal Larmor radius, and assuming a nearly isotropic particle distribution, where, on average, $\langle v_{\perp} \rangle  \approx \langle \left| v_{||} \right| \rangle$, we can further reduce this to $k^{\mathrm{res}} = \langle | k_{||}^{\mathrm{res}} | \rangle \sim r_L^{-1}$. Eq. \ref{Eq:where_resonanted} should therefore be used with care, but can still provide a useful first-order estimation.\\

Two interesting effects are evident from Fig. \ref{fig:spectra_resonate}. Firstly, electrons with $E<100$ keV will resonate in the so-called dissipation range of the slab spectrum. Here, most of the turbulent energy is already dissipated and the turbulence is relatively weak. We will show in the next section that this leads to very little pitch-angle scattering (and a correspondingly large parallel mean free path) at the lowest energies. Secondly, for the 2D component, we note that, even at the highest energies considered here, 

\begin{equation}
k_{\mathrm{res}} \sim r_L^{-1} > \langle \l_{\perp} \rangle^{-1} 
\end{equation}

where $\langle \l_{\perp} \rangle$ is the assumed perpendicular correlation length (shown on the figure as the dashed vertical green line). Assuming that the 2D component is primarily responsible for perpendicular scattering, this means that resonant interactions are most likely negligible and perpendicular transport is governed by the large scale magnetic fluctuations. This allows us to work in the non-resonant so-called field line random walk (FLRW) limit of perpendicular diffusion \citep[][]{straussetal2016}, which should be a fair approximation in the regime $r_L \ll \langle \l_{\perp} \rangle$.\\

Here we employ the transport coefficients as used by \cite{straussetal2017} and the interested reader is referred to that work for a detailed description. From the calculated $D_{\mu \mu}$ and $D_{\perp}$, the corresponding parallel ($\lambda_{||}$) and perpendicular ($\lambda_{\perp}$) mean free paths are calculated, and these are shown in Fig. \ref{fig:transport_coeffs} as a function radial distance (at 100 keV) and energy (at 0.3 au and 1.0 au). The radial dependence of these parameters are at 100 keV, of course, identical to those of \cite{straussetal2017}. We note from the left panel of the figure that $\lambda_{||} \gg L$, with $L$ the focusing length, close to the Sun. This implies that parallel transport near the Sun is expected to be focused-dominated; an effect illustrated later on. The energy dependence of $\lambda_{\perp}$ follows the $D_{\perp} \sim v$ dependence of the FLRW coefficients. The energy dependence of $\lambda_{||}$ is, however, more interesting; below $E\sim100$ keV, $\lambda_{||}$ increases with decreasing energy. This dependence is due to energetic electrons resonating with the dissipation range of the slab turbulence power spectrum and weak particle-wave interactions are indeed expected. This behaviour is predicted from theory \citep[e.g.][]{ts2003}, but has also been confirmed from observations \cite[][]{droge2000}. The sensitivity of electron transport to the assumed shape of the slab spectrum at these high frequencies has also been studied in the past by \citet{EB2013} for low energy Jovian electrons. The fact that $\lambda_{||} > L$ at these low energies will again manifest in focused-dominated transport at low energies. Once again, the effect thereof will be illustrated in the next section.

\section{Simulation Results}

\begin{figure*}
\begin{center}
\includegraphics[width=100mm]{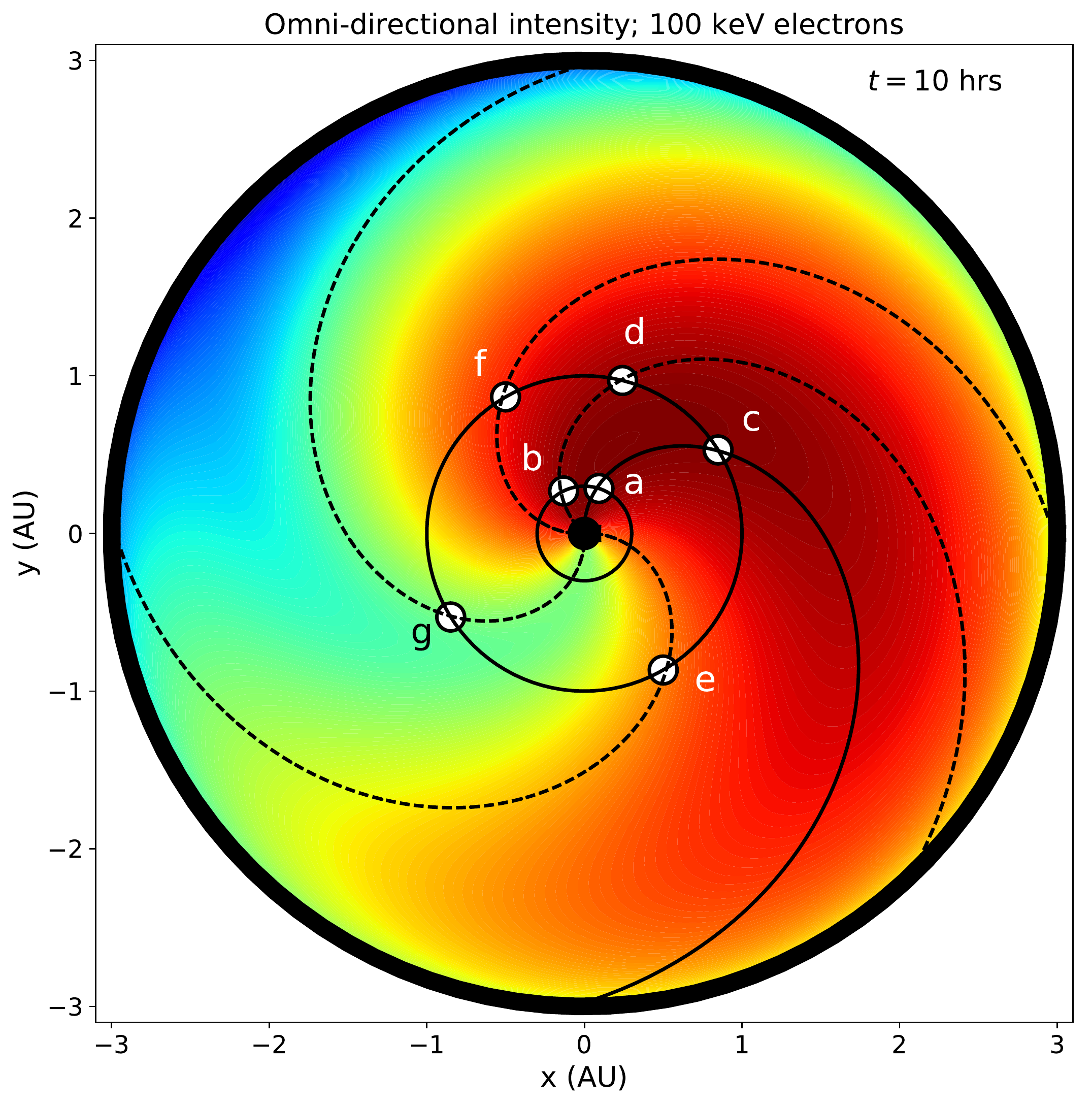}
\caption{The simulated omni-directional intensity of 100 keV electrons in the ecliptic plane for 100 keV electrons at $t=10$ hrs. Also indicated are the position of the 7 virtual spacecraft at which the resulting particle distribution will be investigated further. \label{fig:internsity_contour}}
\end{center}
\end{figure*}

\begin{figure*}
\begin{center}
\includegraphics[width=120mm]{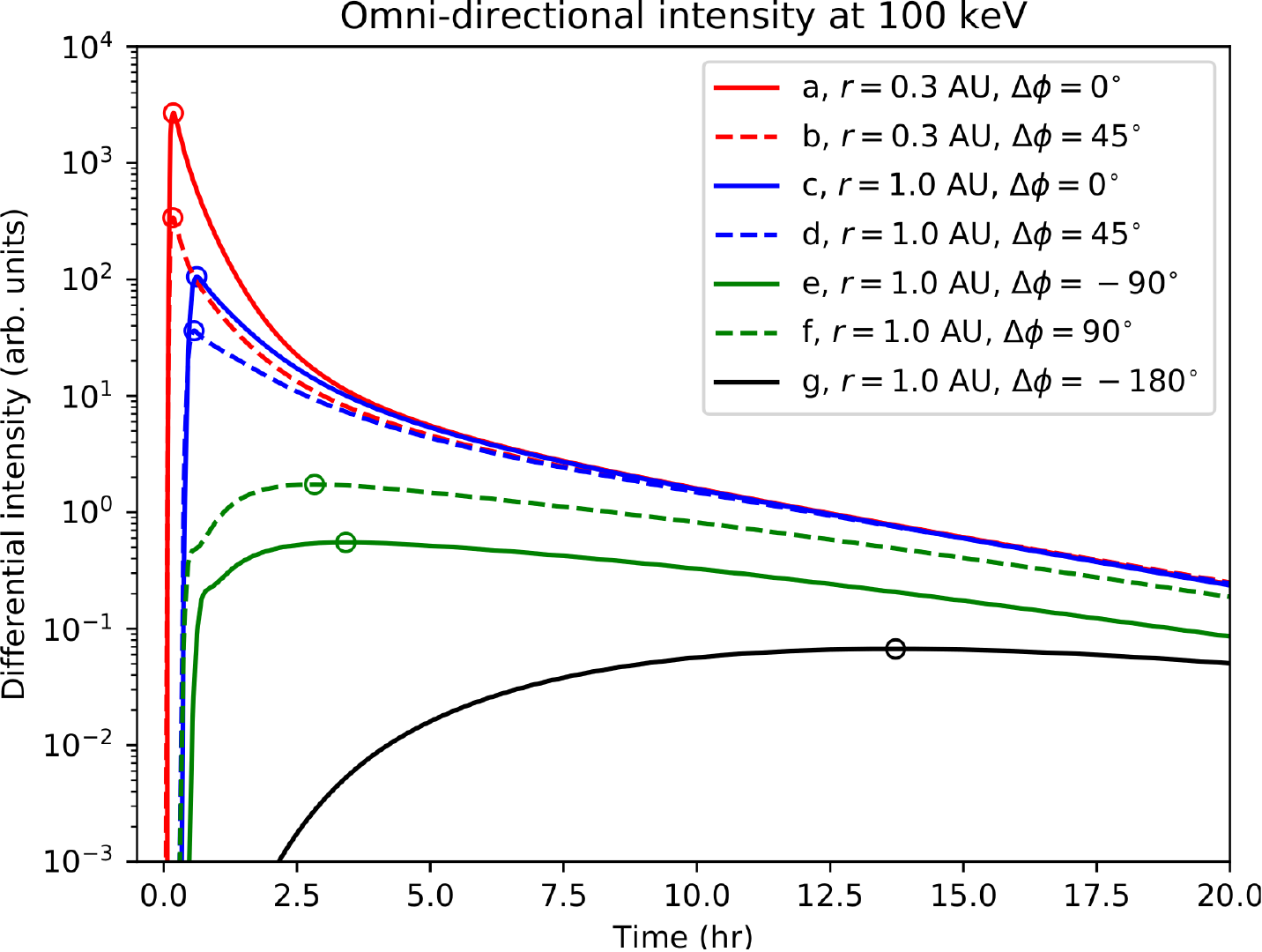}
\caption{The temporal evolution of the omni-directional intensity of 100 keV electrons at the position of the 7 virtual spacecraft. The time-of-maximum (TOM) intensity for each observer is circled.  \label{fig:omni_time}}
\end{center}
\end{figure*}

At the inner boundary of the model, we inject an energy dependent boundary condition, scaling as $E^{-3}$. We follow the usual assumption of specifying the time dependence of this injection function through a Reid-Axford \citep[][]{reid1964} injection profile with $\tau_a=0.1$ hrs and $\tau_e=0.5$ hrs being the assumed acceleration and escape times, respectively. A Gaussian profile in longitude is injected with $\sigma = 5^{\circ}$ \citep[for more details, see again][]{straussetal2017}.\\

The model is solved time-dependently to produce the electron distribution function, for each energy, from which e.g., the omni-directional intensity can be calculated. This quantity is shown, as contour plot for 100 keV electrons, in Fig. \ref{fig:internsity_contour} for $t=10$ hrs. Also shown on this figure are the position of 7 so-called virtual spacecraft (labelled a -- g). At each of these positions, we will examine the time profile and energy dependence of the resulting intensities. These virtual observers are chosen such that: a and c are magnetically connected to the maximum of the injection function (a is at a radial distance of 0.3 au and c at 1 au), b and d are $45^{\circ}$ west of the best magnetic connection at 0.3 and 1 au, respectively, but still relatively well connected to the source, e and f are at 1 au and $90^{\circ}$ east/west of the source and therefore not magnetically connected (in this model, particles will have to diffuse in longitude to reach these observers), and g is at 1 au and $180^{\circ}$ away from source. \\

At each of the virtual spacecraft we start by examining the temporal profile of the omni-directional intensity. This is shown in Fig. \ref{fig:omni_time} for 100 keV electrons. In order to compare the model results with observations, we also calculate the maximum intensity, which are indicated as the open circles on Fig. \ref{fig:omni_time}, at each virtual spacecraft. This is repeated for different energies and a TOM spectrum can be constructed at each spacecraft position.

\subsection{Omni-directional Spectra}

\begin{figure*}
\begin{center}
\includegraphics[width=120mm]{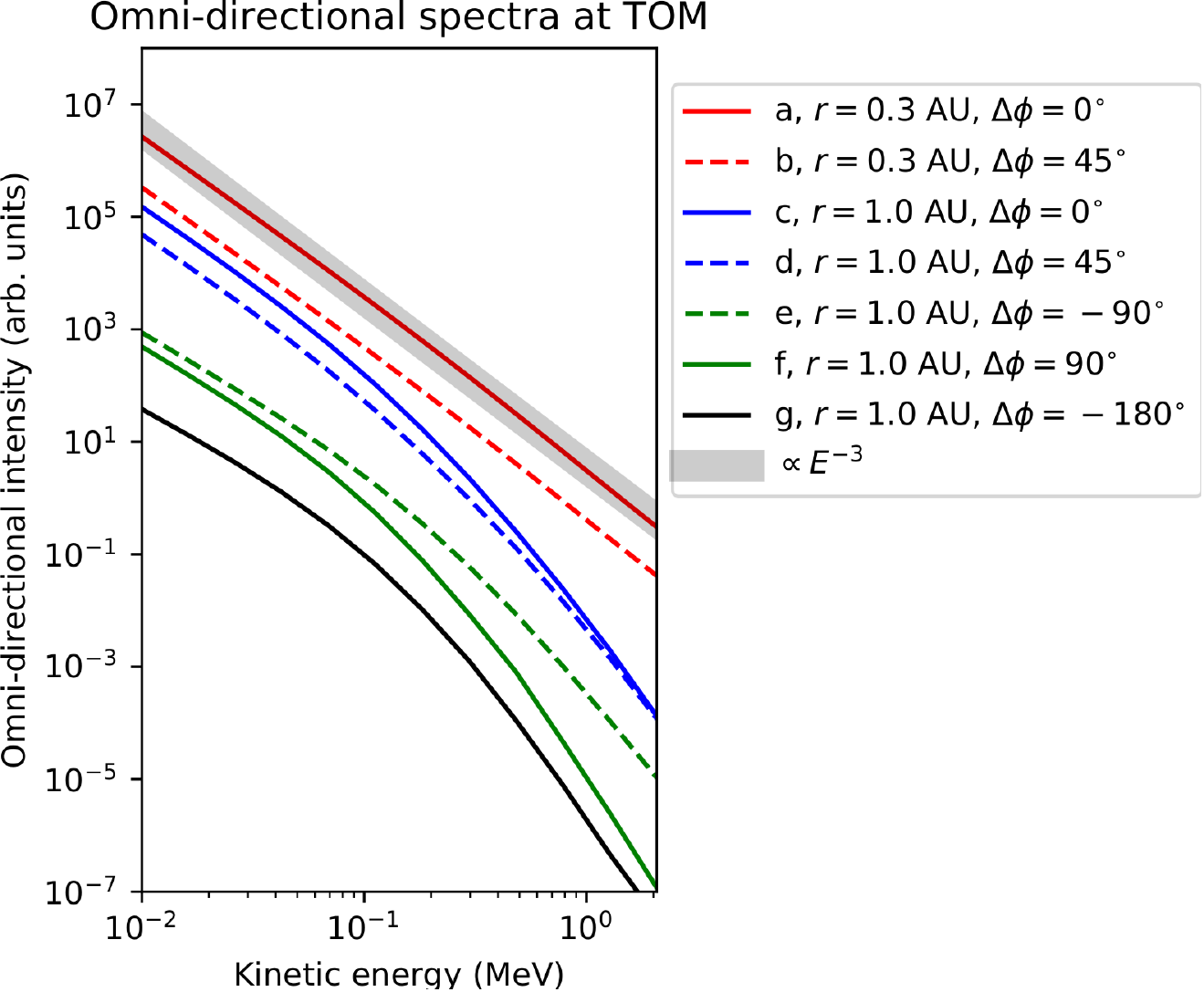}
\caption{The TOM spectrum, calculated for the omni-directional intensity. The different lines correspond to the 7 virtual observers employed in this study, while the grey band illustrates the spectral shape of the injected spectrum. \label{fig:omni_spectra}}
\end{center}
\end{figure*}

Fig. \ref{fig:omni_spectra} shows the TOM spectra for the omni-directional intensity as calculated for the different virtual spacecraft. The grey band shows the energy dependence of the injected spectrum, $E^{-3}$. Interestingly, for the observers at 0.3 au the spectra follow the injected $E^{-3}$ spectral shape. This is a direct consequence of the focused-dominated transport alluded to earlier; particles at these positions suffer very little scattering and, in the absence of any other energy loss/gain processes, the original spectral shape is preserved. The same is true for $E<100$ keV at all other spacecraft positions. We can therefore conclude that when the transport of SEPs is dominated by focusing, the original injected spectral index is preserved. At 1 au and $E>100$ keV, the spectra transition into a softer form due to scattering; when particle scattering becomes significant, particles become more uniformly distributed in space (i.e. more uniform along magnetic field lines) with the result that the TOM intensity decreases, leading to a softer TOM energy spectrum. It is important to note that spectral changes, as presented here, do not imply particle losses (the model presented here conserves the total particle number). \\

\subsection{Pitch-angle Dependent Spectra}

\begin{figure*}
\begin{center}
\includegraphics[height=90mm]{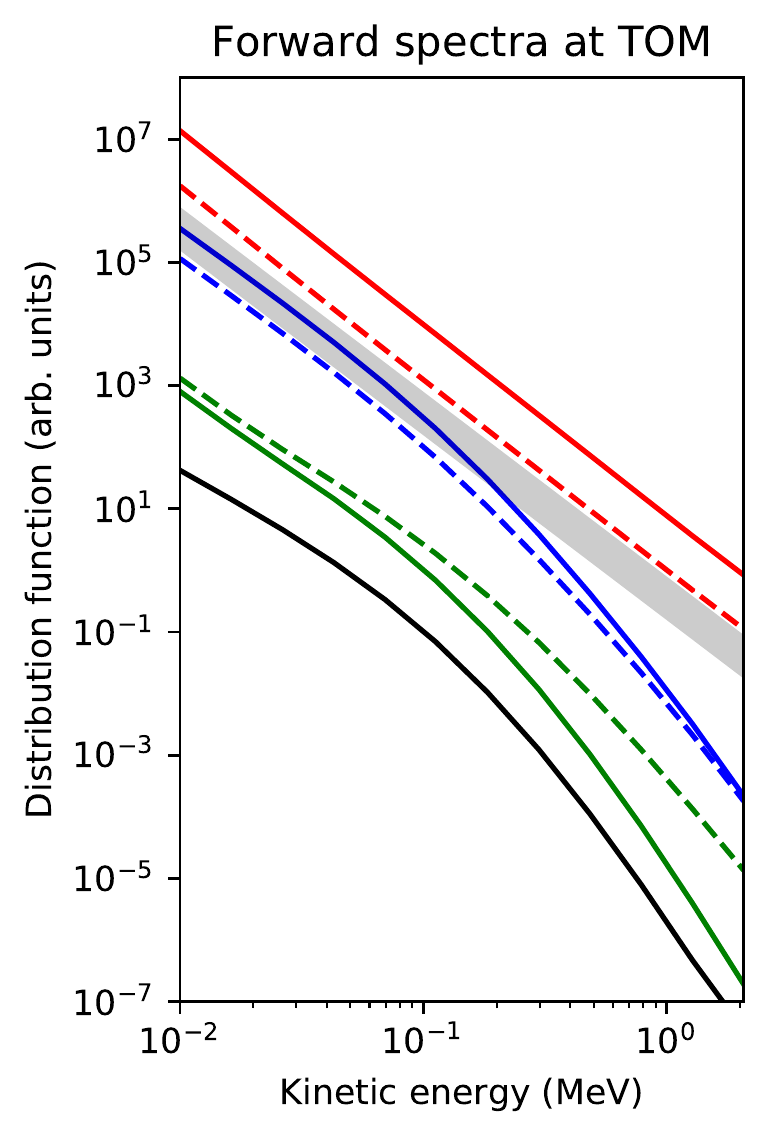}
\includegraphics[height=90mm]{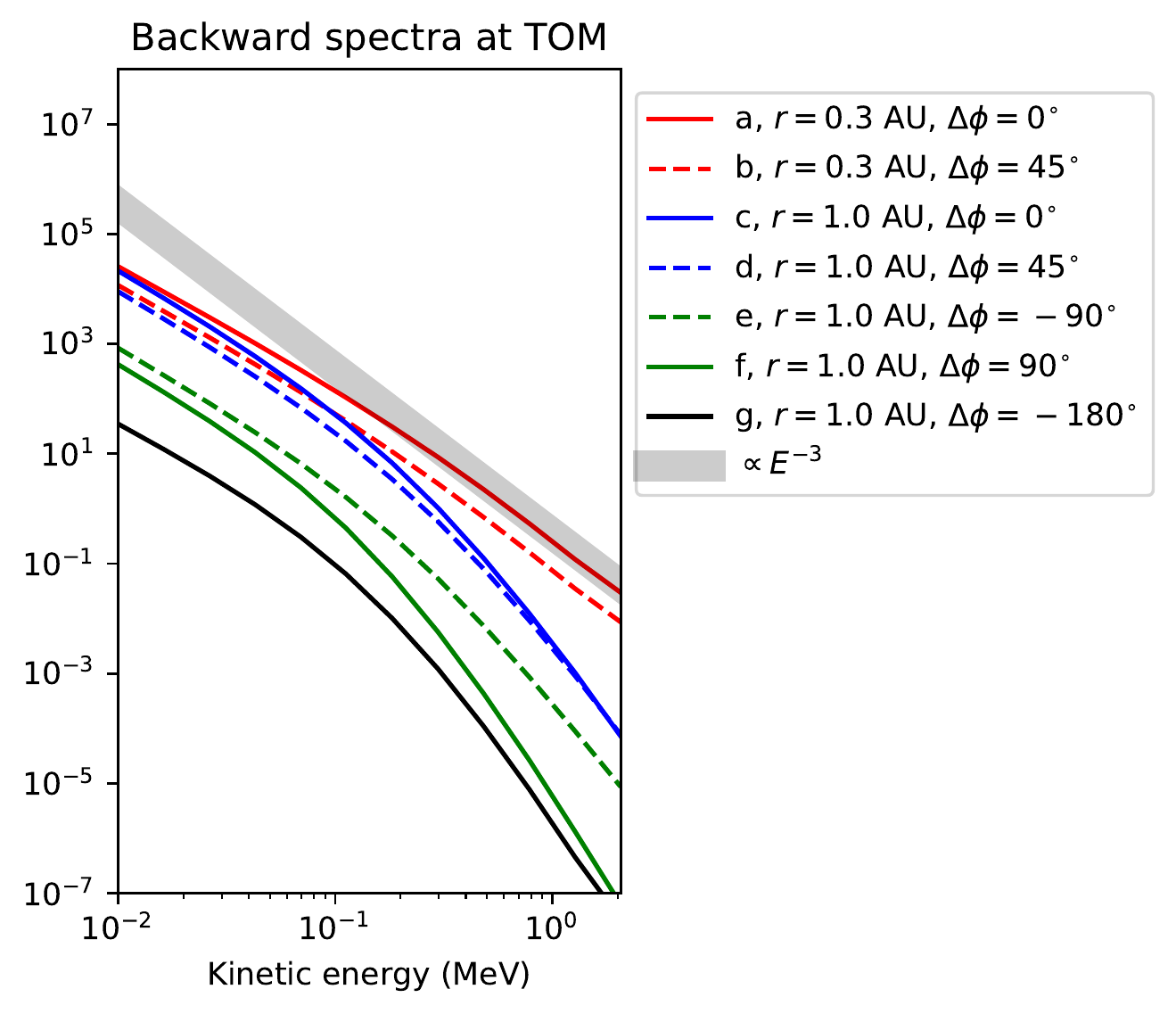}
\caption{Similar to Fig. \ref{fig:omni_spectra}, but now the TOM spectrum is calculated for forward ($\mu = 0.9$, left panel) and backward ($\mu=-0.9$, right panel) propagating electrons separately. \label{fig:spectra_directional}}
\end{center}
\end{figure*}


To test the hypothesis that pitch-angle scattering can significantly affect relativistic SEP electron transport, and can therefore lead to spectral breaks, as well as to compare with the observations presented in section \ref{sec:obs}, we also calculate the TOM spectra for particles with different pitch-angles. For this example we use forward ($\mu = 0.9$) and backward ($\mu = -0.9$) propagating particles. These spectra are constructed for each observer and shown in Fig. \ref{fig:spectra_directional}. The TOM spectra of the $\mu = 0.9$ particles (forward direction) follow that of the omni-directional intensity. This is not surprising as the TOM intensity usually corresponds with the first-arriving particles in a highly anisotropic beam. What is interesting is the spectral hardening of the $\mu = -0.9$ (backside) TOM spectra below $E\sim 100$ keV. At higher energies, where particle scattering becomes significant, both directional spectra have the same shape. However, at energies where the transport is focused-dominated, there is a deficiency of backward propagating particles (perhaps as should be expected), manifesting in a harder spectrum for $\mu = -0.9$ than $\mu = 0.9$ particles. \\


\subsection{Sensitivity to Dissipation Range Onset}

\begin{figure*}
\begin{center}
\includegraphics[width=70mm]{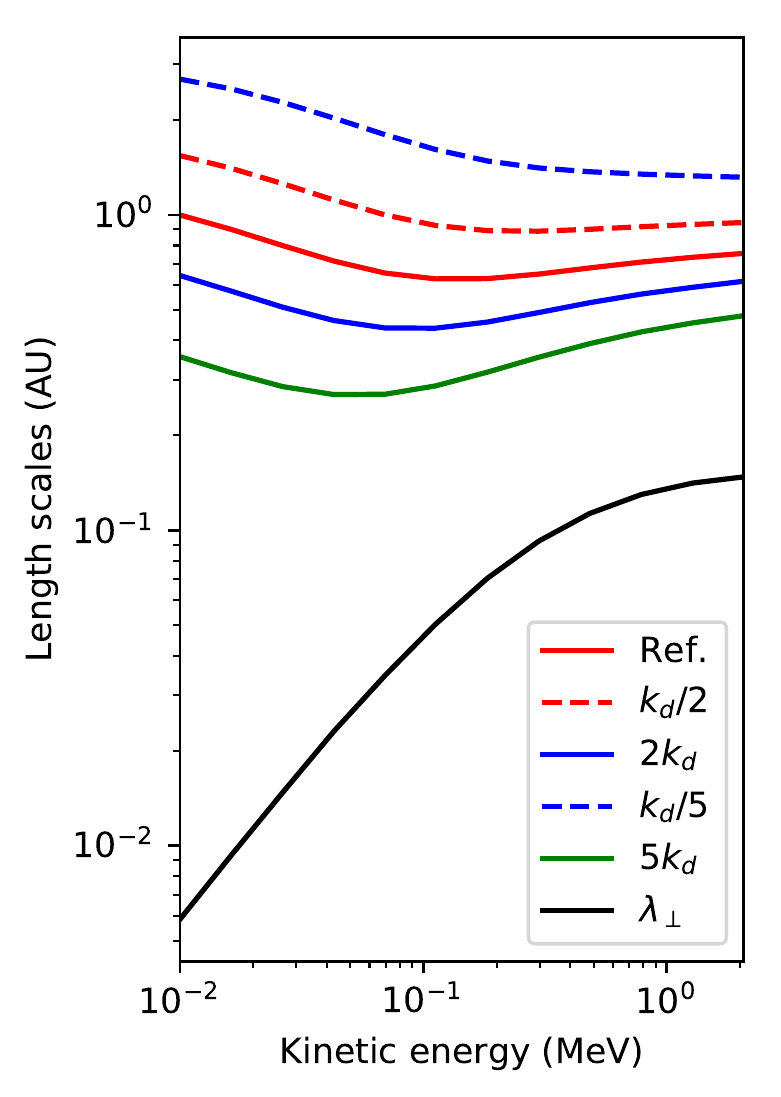}
\hspace{5mm}
\includegraphics[width=70mm]{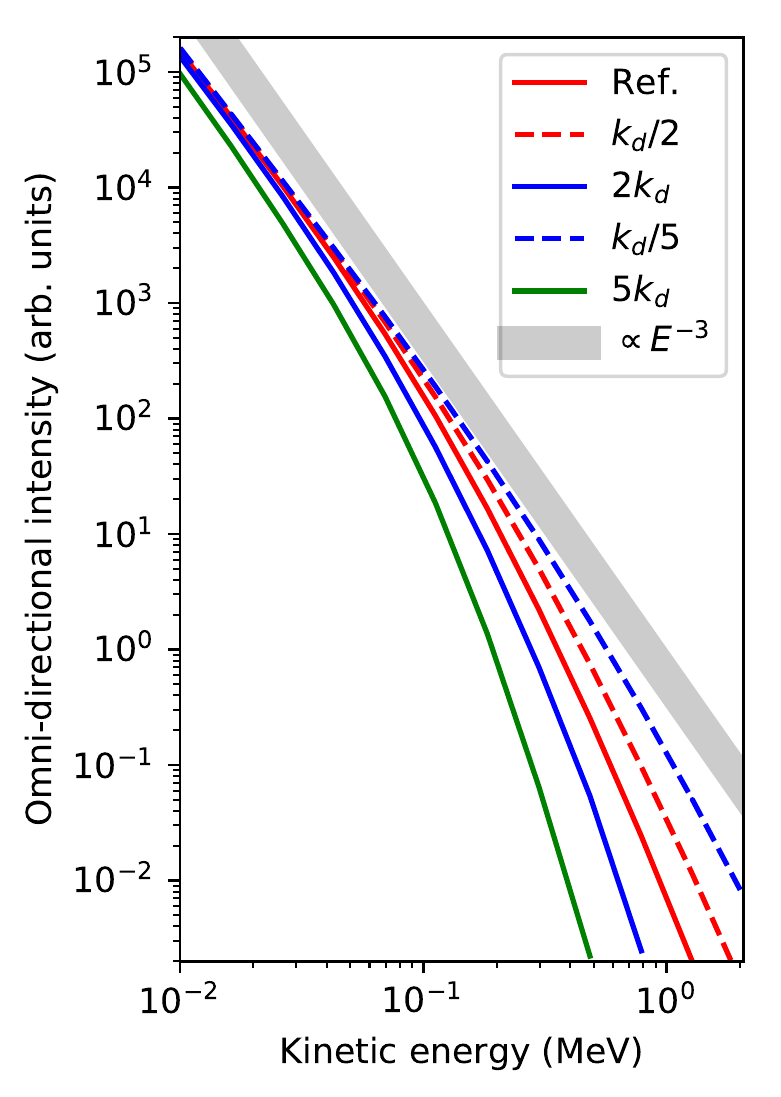}
\caption{The left panel shows the calculated parallel mean-free-paths for various estimates of $k_d$ and the perpendicular mean free path in black. The right panel shows the resulting omni-directional TOM spectra at Earth for an observer magnetically connected to the source.  \label{fig:spectra_different_kd}}
\end{center}
\end{figure*}

It is generally assumed that the dissipation of the wave-like (slab) turbulence component is due to cyclotron damping by the thermal plasma. In this work we characterize this process by specifying the so-called dissipation range onset wave number $k_d$. For this work we use the ``best fit" estimate from \citet{leamonetal2000}

\begin{equation}
\label{Eq:kd}
k_d =  \frac{2 \pi}{V_{sw}} \left( a + b \Omega_i \right),
\end{equation}

where $V_{sw}$ is the solar wind speed, $\Omega_i$ the proton cyclotron frequency, and $a,b = 0.2 \mathrm{Hz},1.76$ are fitting parameters to measurements. Observations by e.g. \citet{smithetal2012} however show that $k_d$ may vary by up to a factor of $\sim 5$ between observations, while modelling of these events suggest changes by a factor of $\sim 2$ can be explained just by changing plasma conditions \citep[see][]{straussengelbrecht2018}. Here we show the sensitivity of both the calculated parallel mean-free-path, and the resulting SEP spectra at 1 au, to different assumed values of $k_d$. We vary the value of $k_d$ by simply multiplying Eq. \ref{Eq:kd} with different factors in a rather ad-hoc fashion. \\

The resulting parallel mean-free-paths corresponding to different choices of $k_d$ is shown in the left panel of Fig. \ref{fig:spectra_different_kd}. For larger values of $k_d$, the inertial turbulence spectrum extends to higher wave numbers and particles experience more scattering. We thus see a decrease in $\lambda_{||}$ with an increase in $k_d$. The right panel of Fig. \ref{fig:spectra_different_kd} shows simulated omni-directional SEP electron spectra for these different choices of $k_d$ for an observer at Earth, magnetically connected to the SEP source. The spectral change from lower to higher energies becomes stronger for larger $k_d$, i.e. smaller $\lambda_{||}$. When $\lambda_{||}$ becomes larger (smaller $k_d$), the break in the spectrum occurs at higher energies; again indicating less scattering.

\section{Discussion}

We find that, on average, anisotropic solar energetic electron events show a spectral hardening over the whole energy range of SEPT (45-425\,keV) when comparing the anti-sunward propagating distribution with the back-scattered distribution. There is a strong event-to-event variation with some events even showing spectral softening. In cases where spectral hardening is observed, most of the events do not show a larger spectral index change than 1.5. However, occasionally changes larger than 2 are observed. Both single-power-law as well as broken-power-law events show the effect. Because the spectral shape often changes from the forward to backward direction, making it difficult to quantify the change, we determine the spectral index change at two representative energies, 70 and 200 keV represented by $\delta70$ and $\delta200$ regardless of the overall spectral shape. In those events which show a broken-power-law shape in both the forward and backward spectrum, we investigated also the change of the break point energy but did not find a systematic effect. The sharpness of the spectral change is yet to be investigated in detail, but preliminary results are presented in Appendix \ref{Sec:sharpness}, indicating that these so-called spectral breaks are, in fact, smoother than initially thought.\\

Using SEP electrons as test particles {in transport simulations} we show that particle transport can lead to changes in the calculated TOM spectrum when particle scattering becomes efficient. For the omni-directional intensity this manifests as a spectral change at $\geq 100$ keV at Earth, although this ``break energy" is highly parameter dependent The ``sharpness" of the modelled spectral change was compared to that of the observations, and the results presented in Appendix \ref{Sec:sharpness}. {These results indicate that the modeled spectral breaks are similar to the roundest (small $\alpha$) observed breaks in the sample or even somewhat rounder. The smoothness of the simulated transition can potentially be influenced by a number of model parameters, including the form of the assumed turbulence dissipation range and the associated resonance function used to determine $D_{\mu \mu}$, or it can simply be a numerical artefact due to calculating physical quantities on a numerical grid. Whatever the reason for this minor discrepancy, we firmly believe that, qualitatively, the simulation results are consistent with the observations presented here.} Traditionally, such observed spectral changes are attributed to the standard theory of diffusive shock acceleration, or some effect from adiabatic energy losses \citep[see e.g.][]{prinslooetal2019}. For SEP electrons this is not the case; the spectral break is rather related to pitch-angle diffusion. In regions where the parallel transport becomes dominated by focusing (at low energies and close to the Sun), the injected spectral form is preserved. {At high energies and larger radial distances, particle scattering becomes efficient and this manifests in a spectral break of the TOM spectrum.} The fact that low energy electrons suffer very little pitch-angle scattering was also recently confirmed, from observational grounds, by \citet{drogeetal2018}. {Furthermore, our simulation results provide an explanation for the unexpectedly high break energies of about 120\,keV found by \citet{Dresing2020}.}\\

The effects discussed above also lead to a low energy spectral hardening of the TOM spectra of backward propagating particles; pitch-angle diffusion is not efficient enough to produce a significant backward propagating particle population. {Although the observations and model results agree in the overall effect of a spectral hardening in the spectrum from forward to backward direction, the observations seem to suggest a stronger effect at higher energies than at low energies which is not understood yet. Potentially, this could indicate that the assumed energy dependence of $D_{\mu \mu}$ still needs further refinement.} {However, the observations may suffer from a number of other effects like potentially time-extended and/or energy-dependent injection functions which influence the forward/backward spectral comparison, and these effects are not presently included in the model}.\\

These simulation results are, of course, sensitive to the choice of assumed parameters. Specifically, for the work presented here, the results are very sensitive to the choice of the dissipation range onset, characterized by $k_d$. We show that relatively small changes in the assumed value of $k_d$ can lead to large changes in $\lambda_{||}$. A similar conclusion was reached by \citet{EB2013} while studying the transport of low energy Jovian electrons. Moreover, \citet{droge2000} shows a large inter-event variability of $\lambda_{||}$ when this quantity is derived from observations. Potentially, this can be explained by changing turbulence conditions, as illustrated here. When we include these different $\lambda_{||}$ scenarios into the numerical model, we find, unsurprisingly, that the simulated SEP spectra change considerably. For larger $\lambda_{||}$ values, the energy break moves to higher energies. These simulations, which now include possible variations in the background turbulence levels, can therefore explain the observed variability in the break energy as derived from SEP measurements. \\

There seems to be only a small longitudinal effect suggesting that perpendicular diffusion, as implemented here, plays little or no role in producing these spectral breaks.\\

Our results indicate that when in-situ SEP electron spectra are compared to remote-sensing observations, as presented by e.g., \citet{kruckeretal2007}, the lower part of the electron spectrum between $\sim 50$ keV and $\sim 100$ keV should be used as this energy regime is least sensitive to transport effects at Earth; below $\sim 50$ keV electrons lose significant amounts of energy due to wave generation mechanisms, while at energies greater than $\sim 100$ keV, pitch-angle scattering can significantly alter the spectral shape.\\

{Simulations suggest that the shape of the omni-directional TOM electron spectra, above $\sim 10 - 60$ keV where wave generation effects becomes negligible, are affected very little by transport close to the Sun (within 0.3 au). This means that {\it Solar Orbiter} and {\it Parker Solar Probe} electron measurements in this higher-energy region will represent the unmodulated SEP electron spectrum representative of the accelerated spectra without the need to account for any transport effects.}

\acknowledgments

We acknowledge helpful discussions and support from the International Space Science Institute through the ISSI team on ``Solar flare acceleration signatures and their connection to solar energetic particles". This work is based on the research supported in part by the National Research Foundation of South Africa (NRF grant numbers 120847, 120345, and 119424). Opinions expressed and conclusions arrived at are those of the authors and are not necessarily to be attributed to the NRF. N.D. was supported under grant 50OC1302 by the Federal Ministry of Economics and Technology on the basis of a decision by the German Bundestag and by the Deutsche Forschungsgemeinschaft (DFG) via project  GI 1352/1-1. N.D. acknowledges support from NASA grant NNX17AK25G. We appreciate funding from the Alexander von Humboldt foundation. The responsibility of the contents of this work is with the authors. Figures were prepared with Matplotlib \citep[][]{hunter}.

\appendix{}

\section{On the sharpness of the spectral changes}
\label{Sec:sharpness}
\begin{figure*}
\begin{center}
\includegraphics[width=0.32\textwidth]{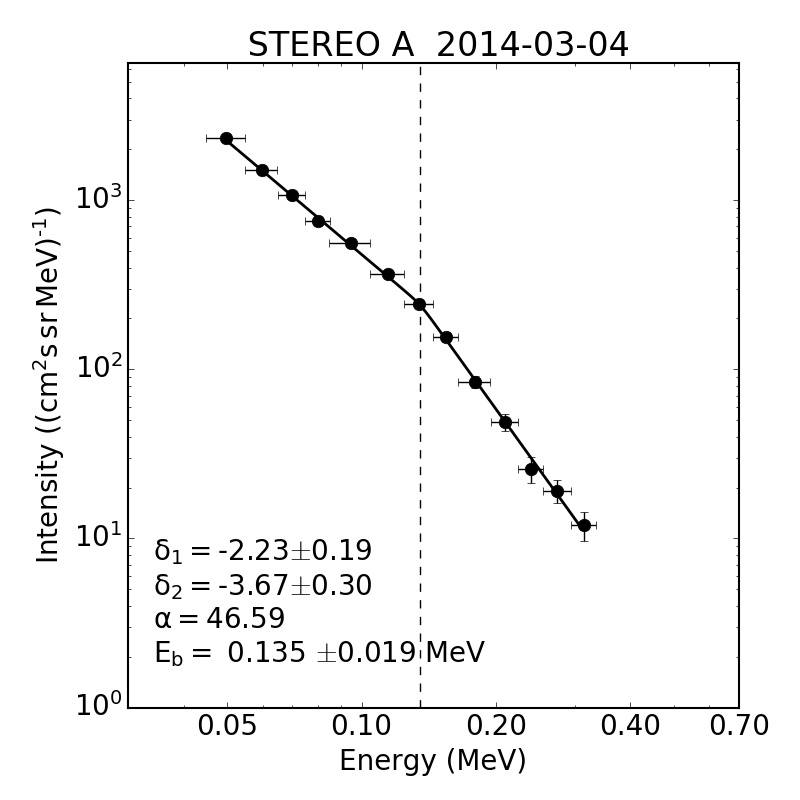}
\includegraphics[width=0.32\textwidth]{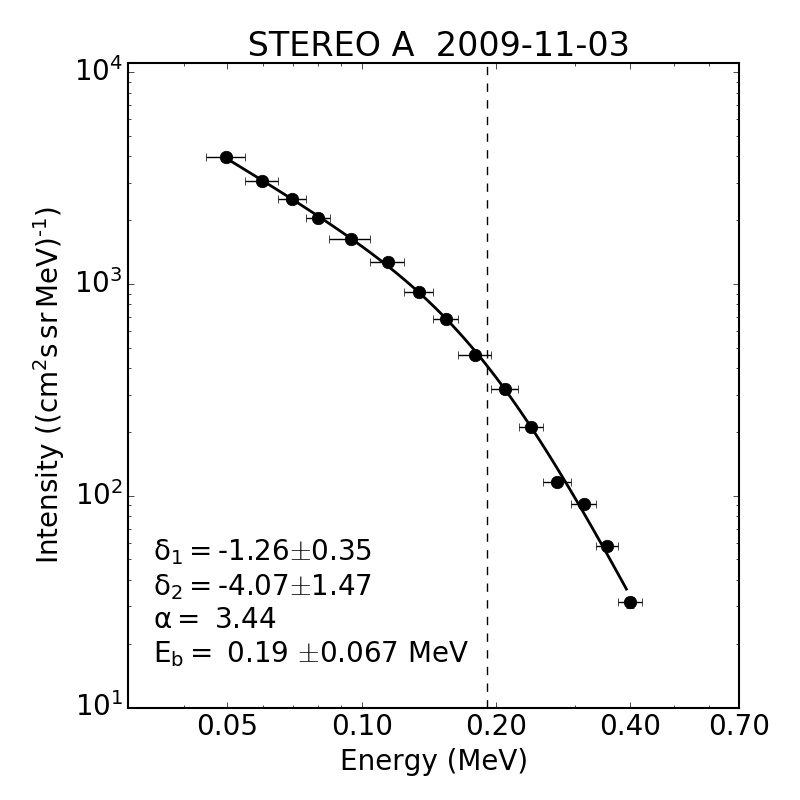}
\includegraphics[width=0.32\textwidth]{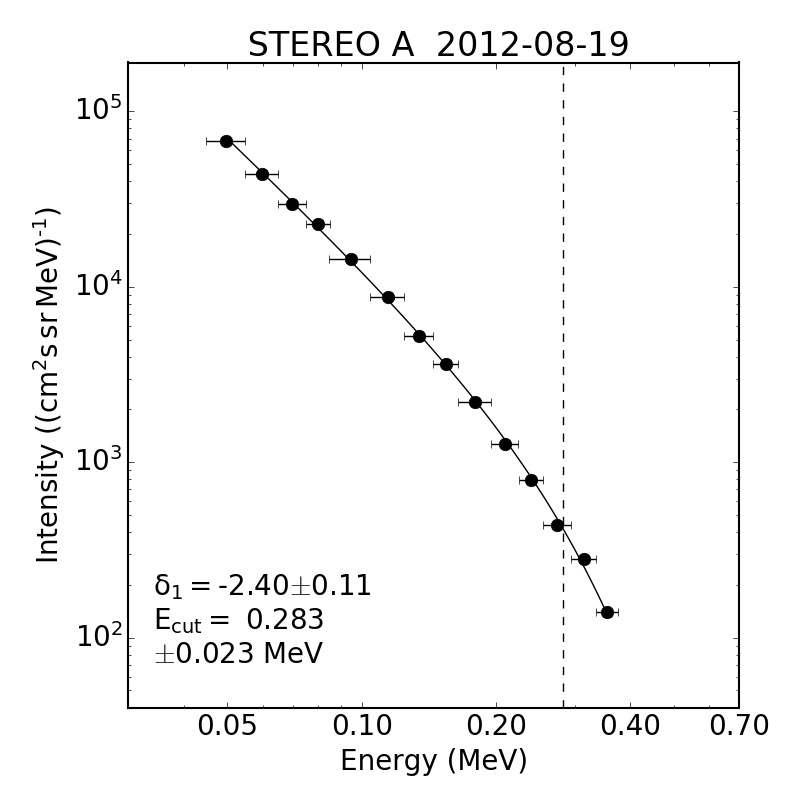}
\caption{{Examples of different spectral shapes observed for near-relativistic electrons with STEREO/SEPT (points). The left and middle spectra are best described by double-power-law functions (solid lines) with the parameter $\alpha$ describing the sharpness of the break (Eq. \ref{Eq:broken_pl}). A larger $\alpha$ yields a sharper break (left) while a smaller $\alpha$ produces a rounder break (middle). The right hand panel shows a spectrum which is best described by a power-law with exponential cutoff (Eq. \ref{Eq:pl_exp}). The figure legends provide the fitting parameters.}} \label{fig:example_spectra}
\end{center}
\end{figure*}
{The large statistical sample of near-relativistic electron spectra which were observed by STEREO/SEPT and build the base of this study show a variety of spectral shapes. \citet{Dresing2020}, who first investigated this sample, showed that about half of the events show only a single-power-law shape while the others resemble a broken or double-power-law. The authors also pointed out that the energy range of the instrument as well as the maximum observed energy of each event can influence the apparent spectral shape. In the present study we fit the observed spectra, taking into account the sharpness of the spectral transition, or break, as described by the parameter $\alpha$ in Eq. \ref{Eq:broken_pl}. In fact most of the double-power-law events do not show very sharp breaks. Fig. \ref{fig:example_spectra} shows three examples of different spectral shapes present in our sample. Note that only those energy bins showing significant signal to noise ratios are included in the plots. While the left panel shows a rather sharp transition, the middle panel is an example of a rather round spectral change described by a smaller $\alpha$ value (see figure legends for the fitting parameters). Because $\alpha$ does not determine the absolute difference between $\delta_1$ and $\delta_2$, its importance for the functional form vanishes for small spectral changes. Furthermore, the energy binning of the data limits the determination of the sharpness of the transition.}\\

{The right panel of Fig. \ref{fig:example_spectra} shows a case which seems to be best described by a single-power-law with an exponential cutoff (Eq. \ref{Eq:pl_exp}) which is now also included in our fitting scheme. Such a spectral shape is found half as often as a double-power-law shape for all events in our sample. Because of the limited energy range of SEPT, or of the specific event, it is not possible in such cases to distinguish if the real spectrum describes an exponential cutoff or a double-power-law function with a rather high break energy, and a very round transition. However, given the fact that we commonly observe double-power-law functions with round transitions, it is suggested that those exponential cutoffs are rather caused by the fact that the power-law part above the spectral transition is not covered well enough by the observations.}\\

{We applied this same fitting technique to the simulation results presented in Fig. \ref{fig:omni_spectra}, and find that these spectra are always best described by a double-power-law. The transition between the different power-laws are, however, generally smoother that those of the observations, with $\alpha \sim 1$.}


\end{document}